\documentclass[preprint,aps,amsmath]{revtex4}
\usepackage{graphicx}
\usepackage{bm}
\usepackage{amsmath}

\begin{document}
\preprint{}
\title{The long hot summer of the tokamak}
\author{Alexander Kendl}
\affiliation{\small Institute for Ion Physics and Applied Physics, 
University of Innsbruck, A-6020 Innsbruck, Austria\vspace{1cm}}
\begin{abstract}
\vspace{0.5cm}
What have the probability for fine weather in summer and the possibility for a
future use of nuclear fusion as a practically unlimited and clean energy
source got in common? 
The answer is in the particular nature underlying both
physical systems: both the atmosphere and hot magnetized fusion plasmas are
determined by similar processes of structure formation in
quasi-two-dimensional periodic nonlinear dynamical systems. Self-organization
of waves and vortices on small scales in both cases leads to large-scale
flows, which are, depending on conditions, either stable for a long time - or
can break apart intermittently and expel large vortex structures. 
In the case of earth's atmosphere, a potential stabilization of the polar jet
stream over northern Europe by warming in early summer leads to a high probability for
stable hot midsummer weather in central Europe. The efficient utilization of
nuclear fusion in a power plant also depends if a stabilization of such zonal
flows ("H mode") may be sustained by heating of the plasma. 
However, instabilities may ruin by rain European summer holidays ("icelandic
lows"), as well as lead to tempestuous eruptions ("ELMs") of energy and particles from the
edge of a fusion plasma onto the walls of the reactor. In the latter case,
this could cause strong erosion of the wall materials and thus an unefficient
operation of a future fusion power plant. Plasma physicists are - similar to
meteorologists - therefore interested in accurate predictions of these
strongly nonlinear dynamical processes. 

\vspace{1.5cm}

{\sl \noindent This is the preprint version of a manuscript accepted for AIP Conf.
Proc. 1445 (2012): Joint ITER-IAEA-ICTP Advanced Workshop on Fusion and Plasma
Physics (Trieste, Italy, 3.-14.10.2011)}
\end{abstract}
\maketitle

\section{ZONAL FLOWS AND THE H-MODE}

Present fusion experiments and future fusion power plants rely on operation in a
high-confinement H mode \cite{Wagner82,Gohil94,Suttrop97}.
The H mode state is characterised by an edge transport barrier related to a
radial electric field $E_r$ in the outer closed flux surface region, which
induces a perpendicular flow with the $E \times B$ velocity. 
A sheared flow arising from a radial variation $E_r(r)$ is invoked to explain
the transport barrier through a suppression of turbulent transport
\cite{Gohil94,Groebner90,Burrell97}: small-scale turbulent vortices are tilted
or sheared apart by the shear flow vorticity, so that the Reynolds stress
transfers energy from the vortices to the mean flow \cite{Biglari90,Diamond91}. 
A flow shear layer can thus be enhanced by turbulence (which in turn it
suppresses), but can also be caused by neoclassical equilibrium electric fields
\cite{Mccarthy93,Heikkinen00}. 

In contrast to such mean $E \times B$ flows, also fluctuating zonal flows and
their interaction with turbulence play an important role in regulating tokamak
and stellarator edge plasma transport \cite{Shats03}. 
Zonal flows appear through turbulent self organisation of quasi 2-d fluids
like magnetised plasmas \cite{Kendl08}, or in geophysical, planetary and stellar
atmospherical and oceanic dynamics \cite{Diamond05,Fujisawa09,Manz09}.  
Zonal flows may be interpreted as a low-frequency spectral condensate phase of
the turbulence \cite{Hallatschek00,Shats03} that posses the highest symmetry
possible in the given geometry: atmospherical and oceanic zonal jets
are latitudinal, and zonal modes in toroidal fusion plasmas are perpendicular
to the magnetic field on flux surfaces. 
The transition between low (L) confinement to the high (H) confinement mode,
which in general occurs in divertor tokamaks after a threshold heating power
is exceeded, can not yet be explained by any theory with predictive quality or
by any first-principles (turbulence) based numerical simulation \cite{Connor00}.
Some descriptive models regard the transition as a predator-prey type
bifurcation, where the turbulence driven zonal flows in turn suppress and
self-regulate the turbulence and transport \cite{Diamond94}.
Toroidal turbulence simulations indicate that the self-consistent zonal flows indeed
significantly moderate but never completely suppress the turbulence \cite{Scott05}.

Edge localised modes (ELMs) are quasi-periodic eruptions from the edge of H
mode plasmas \cite{Zohm96,Connor98a,Becoulet03}.  The ELM burst leads to enhanced transport
and of heat and particles into the open scrape-off layer field line
region and on to the divertor plates \cite{Kamiya07}.  
Some type (III) of ELMs show similarities to bursts in the flow equilibrium found in global
computations of drift wave turbulence \cite{Scott06,Scott07}, while an other
type (I) is assumed to be related to the ideal ballooning mode instability
\cite{Connor98a} and subsequent enhanced turbulent transport \cite{Kendl10}.

\section{JET STREAMS AND THE SEVEN SLEEPERS}

There is a remarkable analogy between the H mode in tokamaks and the
``Seven Sleepers'' summer weather phenomenon occuring in some regions of central
Europe. In both cases stabilised zonal flows act as transport barriers 
in a quasi two-dimensional system.

Jet streams are strong latitudinally extended and narrow flows in the upper
atmosphere that are determined by Rossby wave interaction similar to drift
wave coupling to plasma zonal flows \cite{Diamond05}. 
The similarity in the underlying quasi two-dimensional fluid dynamics is
expressed in the isomorphism between the Charney-Obukhov equation for
baroclinic dynamics in a rotating system \cite{Charney48,Obukhov49} and the
Hasegawa-Mima equation for turbulence in magnetised plasmas \cite{HM78,DHM11}.
Jet streams have a considerable influence on the weather conditions and short
term weather development, and in turn are affected by the dynamics
of rotating weather fronts \cite{Reiter11}. They also couple to oceanic
streams and oscillations. 
While the jet stream zonal flows often appear undulated and broken, they may stabilise
over a period of a few days, and sometimes, but much more rarely, even
over a few weeks. A prominent example is the possibility for stabilisation of
the northern polar jet stream in summer, which can result in a stable weather period over
central Europa for a few weeks that usually starts in early July. 
This phenomenon is resembled in the southern German ``Siebenschl\"afer''
farmer's weather lore ``{\sl Das Wetter am Siebenschl\"afertag sieben Wochen
  bleiben mag}'': the weather condition around the Seven Sleepers day (June
27) might persist for seven weeks. This weather forecasting day is by lore
supposed to be on the day devoted in Christian calendars to the ``Seven
Sleepers'' of Ephesus patron saint legend \cite{Mahlberg}. 

Indeed a statistically significant meteorological summer singularity is
noted to occur in central Europe with the key date in the first week of July
(around July 7): if the northern jet stream has stabilised at that days,
then with some probability is will remain in its position and stabilise the
general weather situation in central Europe for some time between two and
eight weeks. If the jet stream is located far north (over Scotland and
Scandinavia) then troughs of the Icelandic Low are prevented to propagate into central
Europe, and warm, dry and sunny weather determined by the Azores High will prevail. 
If the jet stream position is located more to the south (across central
Europe) or (unfavourably) meridionally broken, then the Icelandic Low will
carry cool and wet conditions further south to regions ranging from southern
France, Switzerland to southern Germany and further, mostly limited by the
Alpine barrier. Then another formulation of the farmer's lore is appropriate:
``{\sl Ist Siebenschl\"afer nass, regnet's ohne Unterlass}'':  
if it is wet at Seven Sleepers, it will go on raining without cease.

According to the German Meteorological Service, the lore is statistically well
established in southern Germany, where it has originated, and tends to cease
to apply further in the north~\cite{DWD}: In the Munich region in southern
Bavaria and around Innsbruck in Austria the {\sl Siebenschl\"afer} rule has a predictive
quality of up to 80~$\%$, i.e. it works well in around four of five years. 
The predictability decreases to around 60-70~$\%$ further north towards to the
middle of Germany (Frankfurt), and at a latitude corresponding to Hamburg or London
it has shrunk to around only 50~$\%$. Note that this is not the probability for
fine weather, but for persistence of the general weather situation after the
singularity.

While some weather singularities may change or diminish over time (e.g. the
former Alpine regional ``Schafsk\"alte'' cool weather snap in middle of
June appears not to apply any more in recent decades \cite{Meteo}), the
{\sl Siebenschl\"afer} rule seems to be a long term weather phenomenon: the
apparent deviation of the singularity date by 10 days (July 7 instead of
June 27) is generally assumed to correspond to the change from Julian to Gregorian
calendar in the 16th and 17th century. The {\sl Siebenschl\"afer} rule thus
appears to be at least half a millennium old, and still applies.
Similar summer weather lore is known from other regions, like St.~Swithun's
day (July 15) in the British Isles, or Saint-Gervais et Saint-Protais day
(June 19) in France, or St.~Godelieve's day (July 6) in Belgium \cite{Denham1846}. 
The predictability quality of these otherwise similar farmer's rules is however
usually lower than for the {\sl Siebenschl\"afer} rule around south Germany.
There are a number of further weather rules corresponding to established weather
singularities in some regions around the world, although the majority of
farmer's weather rules are generally not trustworthy \cite{Mahlberg}. 

The formation and location of the summer norther polar jet stream is linked to
the warming of the northern atmosphere in spring and early summer.
The zonal flow stabilisation of the summer weather is most noticeable in the
years with a hot and dry period. On most occasions the more stable jet stream dominated
period lasts up to around four weeks \cite{Bissolli91} and is often terminated
by south-western disturbances, but may sometimes last a few weeks longer. The
latest of such a ``quiescent H mode summer'' has occurred in 2003. In more recent years
the rule also often had applied correctly in the southern German and Austrian
area, but has mostly led to a continuation of prevailing unstable conditions
throughout these ``L mode'' summers. Then warm periods of a few days have
often been consecutively interrupted (on a nearly weekly basis) by cooling
thunderstorm fronts. Noteably, the average summer temperature still has risen
over the last years, whereas humidity and rain fall have been increased in
such ``Type III  ELMy H mode'' years, like in 2011.  
The coupling between jet stream activity and global 
warming is in any case highly nonlinear and future developments in this complex
dynamical system are hardly predictable. There appears to be a trend towards a
poleward shift of the jet streams over the last decades: while the
subtropical northern jet stream tends to become weaker, the statistics
suggest some trend towards strengthening of the northern polar stream \cite{Archer08}.

\section{OUTLOOK AND CONCLUSIONS}

The quasi two-dimensional fluid systems of the earth atmospherical dynamics
and of magnetised plasma turbulence feature some remarkable analogies,
starting from the isomorphic description in the most simple (but of course for
practical application too much oversimplified) constituing equations, and
showing some similar large-scale structure formation processes,
like the coupling between eddy motion with zonal flows.

The zonal flows in tokamaks and the atmospheric jet streams are under certain
and in both cases not yet completely understood conditions (controlled
by heating) able to stabilize into long term mean flows that can
act as transport barriers on small scale eddy motions. 

Both the fusion plasmas as well as the weather dynamics are highly
nonlinear and complex dynamical systems, which are not (yet) on the long term
predictable by any first principles based theory or simulation. Experience has
in both cases lead to heuristic rules that allow for some statistical prognosis.
Scaling laws derived from existing tokamak experiments lead to a prediction of
a 70-90~$\%$ chance that the next international fusion experiment ITER will
operate in an H mode state (or more specifically, to reach a fusion efficiency
of $Q_0 \geq 10$) \cite{ITER}. Weather recordings lead to a prediction that
the next summer in Innsbruck will have an 80~$\%$ chance to persist in the same
large weather situation (whatever this will be) as in the first week of
July. This is usually a good basis for planning further summer vacations. And,
altogether, these are good reasons to further continue with fusion plasma
physics research in general, and with our quest to understand turbulence and
predict the H mode in particular. 

\section*{ACKNOWLEDGMENTS}

\noindent In memory of Dr.~Horst Wobig, an outstanding mentor in theoretical plasma
physics who always emphasized to behold the whole picture. 

\medskip

\noindent This work was supported by the Austrian Science Fund (FWF) Y398.

\end{document}